\documentclass[
 aip,
 jmp,
 amsmath,amssymb,
 reprint,
]{revtex4-1}

\usepackage{graphicx}
\usepackage{grffile}
\usepackage{dcolumn}
\usepackage{bm}
\usepackage{multirow}
\usepackage{color}

\begin{document}

\preprint{AIP/123-QED}

\title[Can the evolution of music be analyzed in a quantitative manner?]{Can the evolution of music be analyzed in a quantitative manner?}

\author{Vilson Vieira}
 \homepage{http://automata.cc}
 \email{vilson@void.cc}

\author{Renato Fabbri}
 \homepage{http://www.estudiolivre.org/el-user.php?view\_user=gk}
 \email{renato.fabbri@gmail.com}

\author{Gonzalo Travieso}
  \email{gonzalo@ifsc.usp.br}
  \affiliation{ 
Instituto de F\'isica de S\~ao Carlos, Universidade de S\~ao Paulo (IFSC/USP)
}

\author{Luciano da Fontoura Costa}
  \homepage{http://cyvision.ifsc.usp.br/~luciano/}
  \email{ldfcosta@gmail.com}
  \affiliation{ 
Instituto de F\'isica de S\~ao Carlos, Universidade de S\~ao Paulo (IFSC/USP)
}

\date{\today}

\begin{abstract}

We propose a
methodology
to study music development by
applying multivariate statistics on composers characteristics.
Seven representative composers were considered in terms of
eight main musical features. 
Grades
were assigned to each characteristic and their correlations were
analyzed. 
A bootstrap method was
applied to simulate hundreds of artificial composers
influenced by the seven representatives chosen.
Afterwards we quantify non-numeric relations like dialectics, opposition
and innovation.
Composers differences on style and technique were represented
as geometrical distances in the feature space, making it possible to
quantify, for example, how much Bach and Stockhausen differ from other composers or how
much Beethoven influenced Brahms.
In addition, we compared the results with a prior investigation
on
philosophy~\cite{Fabbri}. Opposition, strong on
philosophy, was not remarkable on music. Supporting an observation already considered by music
theorists, strong influences were identified between
composers by the quantification of dialectics, implying inheritance and suggesting a stronger
master-disciple evolution when compared to the philosophy analysis.
\end{abstract}


\maketitle

\section{\label{sec:level1}Introduction}

In the history of music, composers developed their own styles along a
continuous search for coherence or unity. In the words of Anton
Webern~\cite{Webern}, ``[...] ever since music has been written most great artists
have striven to make this unity ever clearer. Everything that has
happened aims at this [...]''. Along this process we can identify
a constant heritage of style from one composer to another as
a gradual development from its
predecessor, contrasting with the necessity for innovation. Quoting
Lovelock: 
``[...] by experiment that progress is possible; it is the man
with the forward-looking type of mind [...] who forces man out of the
rut of `what was good enough for my father is good enough for me'.''~\cite{Lovelock}.
Thus, development in music follows a dichotomy: while composers aims on
innovation, creating their own styles, their technique is based on the
works of their predecessors, in a master-apprentice tradition.

Other fields like philosophy demonstrate
a well-defined trend when considering innovation: unlike music, the
quest for difference seems to drive philosophical
changes~\cite{Deleuze}. Recently, this observation became more evident with
the application of a quantitative method~\cite{Fabbri} where
multivariate statistics was used to measure non-numeric relations and
to represent the historical development as time-series. More specifically, the method consists of
scoring memorable philosophers based on some relevant
characteristics. The group of philosophers was chosen
based on historical relevance. The
scores assigned to each philosopher characteristics define a state
vector in a feature space. Correlations between these
characteristic vectors were identified and principal component
analysis (PCA) was applied to
represent the philosophical history as a planar space where we could
identify interesting properties. Furthermore, concepts
like dialects can be modeled as mathematical relations between the
philosophical states. Here, we extend that analysis to music.

The application of statistical analysis to
music is not recent. On musicology, statistical methods have been used
to identify many musical characteristics.
Simonton~\cite{Simonton1991829, Simonton1977791} used time-series analysis to measure the creative productivity
of composers based on their music and popularity. Kozbelt~\cite{Kozbelt01012009, Kozbelt01012007} also
analyzed the productivity, but based on the measure of performance
time of the compositions and investigated the relation between
productivity and versatility. More recent works~\cite{Kranenburg2004, Kranenburg2007} use machine-learning
algorithms to recognize musical styles of selected compositions.

Differently from these works, we are not interested in applying
statistical analysis to music but on characterizing composers.
Eight characteristics were described and scored by the authors, based
on the recurrent appearance of these attributes in music pieces.
We chose seven representative composers from different periods of
music history. 
This group was chosen purposely to model their
influence on contemporaries, represented as a group of
``artificial composers'', sampled by a bootstrap
method~\cite{Varian}.
The same statistical method used in
philosophy~\cite{Fabbri} was applied to this set of composers and their
characteristics, allowing us to compare the results from both fields.
The
results present contrasting historical facts, recognized along
the history of music, quantified by application of
distance metrics which allowed us to formalize
concepts like dialectics, innovation and opposition, resulting in
interpretations of music development which are
compatible with perspectives from musicians and
theorists~\cite{Webern, Lovelock}.

\section{Mathematical Description}

A sequence $S$ of $C$ music composers was chosen based on their
relevance at each period of the classical music history.
As done for philosophers~\cite{Fabbri}, the set of $C$ measurements
define a $C$-dimensional space henceforth referred as the \emph{musical space}.  
The characteristic vector $\vec{v}_i$ of each composer $i$ defines a respective
\emph{composer state} in the musical space. For the set of
$C$ composers, we defined the same relations adapted for philosophers~\cite{Fabbri}, 
sumarized in Table \ref{tab:tableRelations}. 

\begin{table}
\caption{\label{tab:tableRelations}Description of mathematical relations defined for each composer $i$, $j$ and $k$ given a set of $C$ composers as a time-sequence $S$.}

\begin{ruledtabular}
\begin{tabular}{ll}

\\ Average state & $\vec{a}_i = \frac{1}{i}\sum_{k=1}^i\vec{v}_k.$ \\ \\

Opposite state & $\vec{r}_i = \vec{v}_i + 2(\vec{a}_i - \vec{v}_i)$ \\ \\

Opposition vector & $\vec{D}_i=\vec{r}_i - \vec{v}_i$ \\ \\

Musical move & $\vec{M}_{i,j} = \vec{v}_j - \vec{v}_i$ \\ \\

Opposition index & $W_{i,j} = \frac{\left< \vec{M}_{i,j}, \vec{D}_i\right>}{||\vec{D}_i||^2}$ \\ \\

Skewness index & $s_{i,j} = \sqrt{\frac{|\vec{v}_i-\vec{v}_j|^2|\vec{a}_i-\vec{v}_i|^2 - [(\vec{v}_i-\vec{v}_j) . (\vec{a}_i-\vec{v}_i)]^2}{|\vec{a}_i-\vec{v}_i|^2}}$ \\ \\

Counter-dialectics \\ index & $d_{i \rightarrow k} = \frac{|\left< \vec{v}_j-\vec{v}_i,\vec{v}_k \right> + \frac{1}{2}\left<\vec{v}_i-\vec{v}_j, \vec{v}_i+\vec{v}_j\right>|}{|\vec{v}_j-\vec{v}_i|}$

\end{tabular}
\end{ruledtabular}
\end{table}

\begin{figure}
        \begin{center}
                \includegraphics[width=0.35\textwidth]{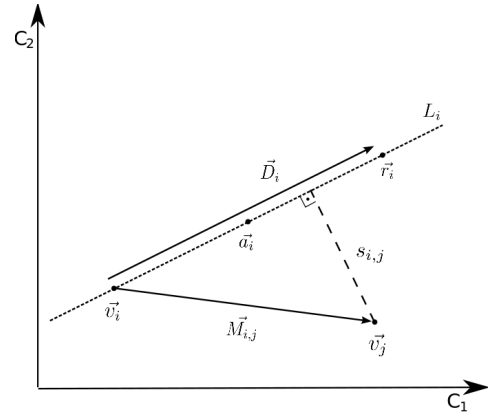}
        \end{center}
        \caption{\it Graphical representation of the measures derived from a \emph{musical move}\cite{Fabbri}.}
        \label{fig.1}
\end{figure}

It is important to note some details about these relations. Given a set of $C$ composers as a time-sequence $S$, the \emph{average state} at time $i$ is defined. The \emph{opposite state} is defined as the ``counterpoint'' of a musical state $\vec{v}_i$, considering its average state: everything running along the opposite direction of $\vec{v}_i$ are understood as opposition. In other words, any displacement from $\vec{v}_i$ along the direction $\vec{r}_i$ is a \emph{contrary move}, and any displacement from $\vec{v}_i$ along the direction $-\vec{r}_i$ is an \emph{emphasis move}. Given a musical state $\vec{v}_i$ and its opposite state $\vec{r}_i$, we can define the \emph{opposition vector} $\vec{D}_i$. These details are better understood analyzing Figure \ref{fig.1}.

Considering the time-sequence $S$ we defined relations between pairs of composers. The \emph{musical move} implied by two successive composers at time $i$ and $j$ corresponds to the $\vec{M}_{i,j}$ vector extending from $\vec{v}_i$ to $\vec{v}_j$. Given the musical move we can quantify the intensity of opposition by the projection of $\vec{M}_{i,j}$ along the opposition vector $\vec{D}_i$, normalized, yelding the \emph{opposition index}. Considering the same musical move, the \emph{skewness index} is the distance between $\vec{v}_j$ and the line $L_i$ defined by the vector $\vec{D}_i$, and therefore quantifies how much the new
musical state departs from the respective opposition move.

\begin{figure}
        \begin{center}
                \includegraphics[width=0.35\textwidth]{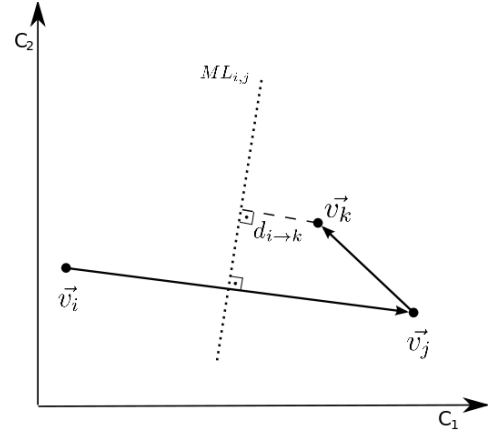}
        \end{center}
        \caption{\it Graphical representation of the quantification of dialectics\cite{Fabbri}.}
        \label{fig.2}
\end{figure}

A relationship between a triple of successive composers can also be defined. Considering $i$, $j$ and $k$ being respectively understood as the \emph{thesis}, \emph{antithesis} and \emph{synthesis}, we defined the \emph{counter-dialectics index} by the distance between the musical state $\vec{v}_k$ and the middle line $ML_{i,j}$ defined by the thesis and antithesis, as shown in Figure \ref{fig.2}. In higher dimensional philosophical spaces,
the middle-hyperplane defined by the points which are at equal
distances to both $\vec{v}_i$ and $\vec{v}_j$ should be used instead
of the middle line $ML_{i,j}$. The proposed equation for counter-dialectics scales to hyperplanes.

The counter-dialectics index is suggested and used instead of dialectics index to maintain compatibility with the use of a distance from point to line as adopted for the definition of skewness.

\section{Musical Characteristics}

To create the musical space we derived eight variables corresponding to
distinct characteristics commonly found in music compositions. The
characteristics are related with the basic elements of music --- melody,
harmony, rhythm, timbre, form and tessitura~\cite{BennettHistory} --- and
non-musical issues like historical events that have influenced the
compositions, for example, the
presence of Church. All the eight
characteristics are listed below:

{\bf \em{ Sacred - Secular}} ($S$-$S_c$): the sacred or religious music is
composed through religious influence or used for its purposes. \textit{Masses},
\textit{motets} and hymns, dedicated to the Christian liturgy, are well known examples~\cite{Lovelock}. Secular
music has no or little relation with religion and includes
popular songs like Italian madrigals and German \textit{lieds}~\cite{BennettHistory}. 

{\bf \em{ Short duration - Long duration}} ($D_s$-$D_l$): compositions are
quantified having short duration when they do not have more than few minutes
of execution. Long duration compositions have at least 20 minutes of execution or
more. The same consideration was adopted by Kozbelt~\cite{Kozbelt01012009,
  Kozbelt01012007} in his analysis of time execution.

{\bf \em{ Harmony - Counterpoint}} ($H$-$C$): harmony regards the
vertical combination of notes, while counterpoint focuses on
horizontal combinations~\cite{BennettHistory}.

{\bf \em{ Vocal - Instrumental}} ($V$-$I$): compositions using just vocals
(e.g.\ \emph{cantata}) or exclusively instruments
(e.g.\ \emph{sonata}). It is interesting to note the use of
vocals over instruments on Sacred compositions~\cite{Lovelock}.

{\bf \em{ Non-discursive - Discursive}} ($D_n$-$D$): compositions
based or not
on verbal discourse, like programmatic music or Baroque rhetoric, where the composer wants
to ``tell a history'' invoking images to the listeners
mind~\cite{BennettHistory}. Its contrary part is known as
\textit{absolute music} where the music is written to be appreciated simply
by what it is.

{\bf \em{ Motivic Stability - Motivic Variety}} ($M_s$-$M_v$): motivic pieces presents equilibrium
between repetition, reuse and variation of melodic motives. Bach is noticeable by his
\textit{development by variation} of motives, contrasting with the
constantly inventive use of new materials by Mozart~\cite{Webern}.

{\bf \em{ Rhythmic Simplicity - Rhythmic Complexity}} ($R_s$-$R_c$): presence or not of polyrhythms, the
use of independent rhythms at the same time --- also known as
\textit{rhythmic counterpoint}\cite{BennettHistory} --- a characteristic
constantly found in Romanticism and the works of 20th-century composers like Stravinsky.

{\bf \em{ Harmonic Stability - Harmonic Variety}} ($H_s$-$H_v$):
rate of tonality change along a piece or its stability. After the highly
polyphonic development in Renaissance, Webern regarded Beethoven as the
composer who returned to the maximum exploration of harmonic variety~\cite{Webern}.

\section{Results and Discussion}
\label{sec:results}

Memorable composers were chosen as key representatives
of musical development. 
This group was chosen purposely to model their influence
over contemporaries, creating a concise parallel with music history. We modeled this group of influenced
composers as new artificial samples generated by a bootstrap method, better
explained in this section.

The sequence
is ordered chronologically and presented on Table \ref{tab:table0} with
each composer related with its historical period.

\begin{table}[ht]
\caption{\label{tab:table0} The sequence of music composers ordered chronologically
with the period each represent.}

\begin{tabular}{|l||l|}
\hline

 Composer       &  Movement \\ \hline

 Monteverdi      & Renaissance \\
 Bach            & Baroque \\
 Mozart          & Classical \\
 Beethoven       & Classical $\to$ Romantic \\
 Brahms          & Romantic \\
 Stravinsky      & 20th-century \\
 Stockhausen     & Contemporary\\

\hline
\end{tabular}
\end{table}

The quantification of the eight musical
characteristics was performed jointly by the authors of this
article and is shown in Table \ref{tab:tableA}. The scores were
numerical values between 1 and 9. Values more close of 1 reveals the
composer tended to the first element of each characteristic pair, and
vice versa. We emphasize that the focus of this work is not on the specific 
characteristics used or their attributed numerical values,
which can be disputed, but on the techniques employed for the quantitative analysis.

\begin{table}[ht]

\caption{\label{tab:tableA}Quantification of the
eight music characteristics for each of the seven composers.}

\begin{ruledtabular}
\begin{tabular}{|l|c|c|c|c|c|c|c|c|}
\footnotesize
\footnotesize Composer    & \tiny  $S$-$S_c$ & \tiny  $D_s$-$D_l$ & \tiny  $H$-$C$ & \tiny  $V$-$I$ & \tiny  $D_n$-$D$ & \tiny  $M_s$-$M_v$ & \tiny  $R_s$-$R_c$ & \tiny  $H_s$-$H_v$  \\
\hline
 \footnotesize Monteverdi   & 3.0 & 8.0 & 5.0 & 3.0 & 7.0 & 5.0 & 3.0 & 7.0  \\
 \footnotesize Bach         & 2.0 & 6.0 & 9.0 & 2.0 & 8.0 & 2.0 & 1.0 & 5.0  \\
 \footnotesize Mozart       & 6.0 & 4.0 & 1.0 & 6.0 & 6.0 & 7.0 & 2.0 & 2.0  \\
 \footnotesize Beethoven    & 7.0 & 8.0 & 2.5 & 8.0 & 5.0 & 4.0 & 4.0 & 7.0  \\
 \footnotesize Brahms       & 6.0 & 6.0 & 4.0 & 7.0 & 4.5 & 6.5 & 5.0 & 7.0  \\
 \footnotesize Stravinsky   & 8.0 & 7.0 & 6.0 & 7.0 & 8.0 & 5.0 & 8.0 & 5.0  \\
 \footnotesize Stockhausen  & 7.0 & 4.0 & 8.0 & 7.0 & 5.0 & 8.0 & 9.0 & 6.0  \\

\end{tabular}
\end{ruledtabular}
\end{table}

This data set defines an 8-dimensional musical space where each dimension
corresponds to a characteristic that aplies to all 7 composers. 
Such small data set is not adequate for statistical analysis and the imediate analysis of this set would
be highly biased by the small sample.

\subsection{Bootstrap method for sampling \emph{artificial composers}}

To simulate a more realistic musical trajectory, we used a bootstrap
method for generating \emph{artificial composers} contemporaries of those seven chosen.

The bootstrap routine generated randomized scores $\vec{r}$. The
values are not totally random, following a probability distribution
that models the original $n = 7$ scores, given by 
$p(\vec{r}) = \sum^n_{i=1} e^{\frac{d_i}{2\sigma^2}}$
where $d_i$ is the distance between a random score $\vec{r}$
and the original score chart. For each step a
value $p(\vec{r})$ is generated and compared with a random normalized value,
characterizing the Monte Carlo~\cite{Robert2011} method to choose a set of samples. This
samples simulates new randomized composers score charts --- while respecting the
historical influence of the main 7 original exponents. Higher
values of $p(\vec{r})$ imply a stronger influence of the original scores
over $\vec{r}$. For the analysis
we used 1000 bootstrap samples obtained by the bootstrap process
together with the original scores,
considering $\sigma = 1.1$. Other values for $\sigma$ were used yelding 
distributions with bootstrap samples closer to or further from the original 
musical states, which does not affected the musical space substantially.

Pearson correlation coefficients between the eight musical
characteristics chosen are presented in Table \ref{tab:tableB}.
Emphasized coefficients have absolute values larger than 0.5.

\begin{table}[ht]
\caption{\label{tab:tableB}Pearson correlation coefficients between
  the eight musical characteristics.}

\begin{ruledtabular}
\begin{tabular}{|c||c|c|c|c|c|c|c|c|}

-    & \tiny  $S$-$S_c$ & \tiny  $D_s$-$D_l$ & \tiny  $H$-$C$ & \tiny  $V$-$I$ & \tiny  $D_n$-$D$ & \tiny  $M_s$-$M_v$ & \tiny  $R_s$-$R_c$ & \tiny  $H_s$-$H_v$  \\ \hline
\tiny$S$-$S_c$ & -     &  -0.2 &  -0.06 &  \textbf{0.69}  & -0.18 &  0.19   &  \textbf{0.56} &  -0.16 \\
\tiny$D_s$-$D_l$ & -     &  -     &  -0.14 &  -0.13          &  0.2  &  -0.48  &  -0.2         &  0.37 \\
\tiny$H$-$C$ & -     &  -     &  -     &  -0.23          &  0.26  &  0.05  &  0.46          &  0.03 \\
\tiny$V$-$I$ & -     &  -     &  -     &  -              & -0.33 &  0.17   &  0.42          &  -0.06 \\
\tiny$D_n$-$D$ & -     &  -     &  -     &  -              &  -    &  -0.3  &  0.02          &  -0.22 \\
\tiny$M_s$-$M_v$ & -     &  -     &  -     &  -              &  -    &  -      &  0.26          &  -0.15 \\
\tiny$R_s$-$R_c$ & -     &  -     &  -     &  -              &  -    &  -      &  -             &  -0.02 \\
\tiny$H_s$-$H_v$ & -     &  -     &  -     &  -              &  -    &  -      &  -             &  - \\

\end{tabular}
\end{ruledtabular}
\end{table}

We can identify some interesting relations between the pairs of
characteristics that reflect important facts in music history. For
instance, the Pearson correlation coefficient of 0.69 was obtained for
the pairs $S$-$S_c$ (Sacred or Secular) and $V$-$I$ (Vocal or Instrumental),
which indicate that sacred music tends to be more vocal than
instrumental. The coefficient of 0.56 also shows it does not commonly use polyrhythms as we can see
analysing the pairs $S$-$S_c$ and $R_s$-$R_c$ (Rhythmic Simplicity or Complexity).
Negative coefficients of -0.33 for the pairs $V$-$I$ and $D_n$-$D$
(Non-discursive or Discursive) indicated that composers who used
just voices on their compositions also preferred to use programmatic
musics techniques like baroque rhetoric.

PCA was applied to this set of data, yielding the new variances given
in Table \ref{tab:tableC} in terms of percentages of total variance.
We can note the concentration of variance along the four
first PCA axes, a common effect also observed while analyzing
philosophers characteristics~\cite{Fabbri}. This would usualy mean that we could
consider just four dimensions but as we will see below our measurements
differs considerably with the inclusion of all eight components.

\begin{table}[ht]
\caption{\label{tab:tableC}New variances after PCA, in percentages for
  scores on \ref{tab:tableB}.}

\begin{tabular}{|c||c|}
\hline
Eigenvalue  & Value     \\ \hline

$\lambda_1$ &  32 \% \\
$\lambda_2$ &  20 \% \\
$\lambda_3$ &  17 \% \\
$\lambda_4$ &  14 \% \\
$\lambda_5$ &   7 \% \\
$\lambda_6$ &   5 \% \\
$\lambda_7$ &   3 \% \\
$\lambda_8$ &   3 \% \\
\hline

\end{tabular}
\end{table}

\subsection{Robustness to perturbation of the original scores}

As done for philosophers analysis, we performed 1000 perturbations of
the original scores by adding to each score the values -2, -1, 0, 1 or 2 with
uniform probability. In other words, we wanted to test if scoring
errors could be sufficient to cause relevant effects
on the PCA projections. Interestingly, the values of average and
standard deviation for both original and perturbed positions listed in Table
\ref{tab:tableD} show relatively small changes. It is therefore
reasonable to say that the small errors in the values assigned as scores of composers
characteristics do not affected too much its quantification.

\begin{table}
\caption{\label{tab:tableD}Averages and standard deviations of the 
deviations for each composer and for the 
8 eigenvalues.}

\begin{tabular}{|c||c|c|}
\hline

Composers & $\mu_{\Delta}$ & $\sigma_{\Delta}$ \\
\hline

Monteverdi     & 3.7347 & 0.8503 \\
Bach           & 5.3561 & 0.9379 \\
Mozart         & 4.4319 & 0.8911 \\
Beethoven      & 3.4987 & 0.7851 \\
Brahms         & 3.0449 & 0.6996 \\
Stravinsky     & 3.6339 & 0.7960 \\
Stockhausen    & 4.2143 & 0.9029 \\
\hline \hline
Eigenvalues & $\mu_{\Delta}$ & $\sigma_{\Delta}$ \\
\hline
$\lambda_1$ &  -0.1759 & 0.0045 \\
$\lambda_2$ &  -0.0638 & 0.0026 \\
$\lambda_3$ &  -0.0411 & 0.0021 \\
$\lambda_4$ &  -0.0144 & 0.0019 \\
$\lambda_5$ &   0.0578 & 0.0021 \\
$\lambda_6$ &   0.0736 & 0.0023 \\
$\lambda_7$ &   0.0080 & 0.0027 \\
$\lambda_8$ &   0.0835 & 0.0030 \\
\hline

\end{tabular}
\end{table}

\subsection{Results}

Table \ref{tab:Deviates} shows the normalized weights
of the contributions of each original property on the eight
axes. Most of the characteristics contribute almost equally
in defining the axes. 

\begin{table}[ht]
\caption{\label{tab:Deviates}Percentages of
the contributions from each musical characteristic on the eight
new main axes.}

\begin{tabular}{|c||c|c|c|c|c|c|c|c|}
\hline
Musical         & \multirow{2}{*}{$C_1$} & \multirow{2}{*}{$C_2$} & \multirow{2}{*}{$C_3$} & \multirow{2}{*}{$C_4$} & \multirow{2}{*}{$C_5$} & \multirow{2}{*}{$C_6$} & \multirow{2}{*}{$C_7$} & \multirow{2}{*}{$C_8$}\\
Charac. & & & & & & & & \\
\hline
 $S$-$S_c$              &  19.78  &   4.04  & 10.38 & 10.60 &  17.55  &  36.60  &  4.41 &  0.63 \\
 $D_s$-$D_l$            &  13.63  &   9.21  & 19.17 &  3.55 &   3.13  &   1.65  & 25.55 & 24.05 \\
 $H$-$C$                &   1.44  &  26.62  & 8.26 & 13.97 &  21.71  &   7.76  & 13.98 & 12.20 \\
 $V$-$I$                &  18.35  &  12.82  & 9.29 &  8.02 &   9.37  &  40.95  &  2.12 &  2.03 \\
 $D_n$-$D$              &   6.31  &  10.73  & 15.48 & 26.29 &  4.04  &   1.86  & 25.29 &  2.35 \\
 $M_s$-$M_v$            &  16.94  &  13.28  & 15.03 &  4.84 &  32.25  &  1.70  &  2.62 &  4.37 \\
 $R_s$-$R_c$            &  14.13  &   3.26  & 15.58 & 13.80 &   7.48  &  1.88  &  1.36 & 35.99 \\
 $H_s$-$H_v$            &   9.38  &  20.00  &  6.75 & 18.88 &   4.45  &  7.56  & 24.62 & 18.36 \\
\hline
\end{tabular}
\end{table}

\begin{figure}[htbp]
  \begin{center}
    \includegraphics[width=0.45\textwidth]{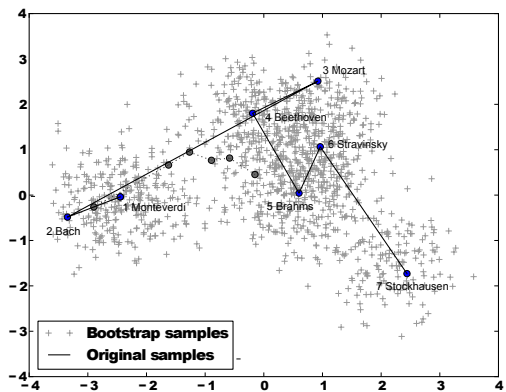}
  \end{center}
  \caption{\it 2-dimensional projected musical space.}
  \label{fig:pca}
\end{figure}

Figure \ref{fig:pca} presents a 2-dimensional space considering the
first two main axes. The arrows follows the time sequence along with the seven
composers. Each of these arrows corresponds to a musical move from one
composer state to another -- for clarity, just the lines of the arrows
are preserved. The bootstrap
samples define clusters around the original
composers.

Bach is found far from the rest of
composers, which suggests his key role
acknowledged by other great composers like Beethoven and
Webern~\cite{Webern}: ``In fact Bach composed everything, concerned
himself with everything that gives food for thought!''. 
The greatest subsequent change takes place from Bach to Mozart,
reflecting a substantial difference in style.
We can identify a strong relationship between
Beethoven and Brahms, supporting the belief by the \textit{virtuosi} Hans von B\"{u}low~\cite{Bulow} when he
stated the $1^{st}$ Symphony of Brahms as, in reality, being the \textit{$10^{th}$ Symphony of
Beethoven}, clamming Brahms as the true successor of
Beethoven. Stravinsky is near to Beethoven and Brahms,
presumably due to his heterogeneity~\cite{BennettHistory,
  Lovelock}. Beethoven is also near to Mozart who deeply influenced
Beethoven, mainly in his early works.
For Webern, Beethoven was the unique classicist who really came close
to the coherence found in the pieces of the Burgundian School: ``Not even
in Haydn and Mozart do we see these two forms as clearly as in
Beethoven. The period and the eight-bar sentence are at their purest
in Beethoven; in his predecessors we find only traces of them''~\cite{Webern}. It
could explain the proximity of Beethoven to the Renaissance  Monteverdi.
Stockhausen is a deviating point when compared with the others and it
could present even more detachment if we had considered
vanguard characteristics --- e.g.\ timbre exploration by using
electronic devices~\cite{Lovelock} --- not
shared by his precursors.

To complement the analysis, Table \ref{tab:tableOI} gives the
opposition and skewness indices for each of the six musical moves,
showing the movements are driven by rather small opposition and strong
skewness. In other words, most musical moves seems to seek more innovation
than opposition. Dialectics is also shown in Table
\ref{tab:tableE} and will play a key role in the next section.

\begin{table}[ht]
\caption{\label{tab:tableOI}Opposition and skewness indices for each
of the six musical moves.}

\begin{tabular}{|c||c|c|}
\hline
Musical Move & $W_{i,j}$ & $s_{i,j}$ \\
\hline \hline

 Monteverdi $\to$ Bach             &   1.0     &  0.      \\
 Bach $\to$ Mozart                 &   1.0196  &  1.9042  \\
 Mozart $\to$ Beethoven            &   0.4991  &  2.8665  \\
 Beethoven $\to$ Brahms            &   0.2669  &  1.7495  \\
 Brahms $\to$ Stravinsky           &   0.4582  &  2.6844  \\
 Stravinsky $\to$ Stockhausen      &   0.2516  &  3.1348  \\

\hline
\end{tabular}
\end{table}

\begin{table}[ht]
\caption{\label{tab:tableE} Counter-dialectics index for each
of the five subsequent pairs of musical moves considering the 8 components.}

\begin{tabular}{|c||c|}
\hline
Musical Triple & $d_{i \rightarrow k}$ \\
\hline \hline

 Monteverdi $\to$ Bach $\to$ Mozart          & 2.0586 \\
 Bach $\to$ Mozart $\to$ Beethoven           & 1.2020 \\
 Mozart $\to$ Beethoven $\to$ Brahms         & 1.0769 \\
 Beethoven $\to$ Brahms $\to$ Stravinsky     & 0.2518 \\
 Brahms $\to$ Stravinsky $\to$ Stockhausen   & 0.2549 \\

\hline
\end{tabular}
\end{table}

\begin{figure}[ht]
        \begin{center}
          \includegraphics[width=0.45\textwidth]{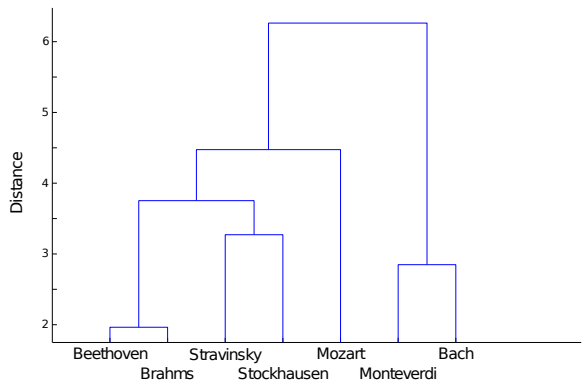}
        \end{center}
        \caption{\it Wards hierarchical clustering of the seven composers.}
        \label{fig:dendrogram}
\end{figure}

We performed Wards hierarchical
clustering~\cite{Ward} to complement the analysis. This algorithm clusters the original scores taking into
account their distance. The generated dendrogram in
Figure \ref{fig:dendrogram} shows the composers
considering their similarity. The representation supports the
observations discussed previously. It is interesting to note the cluster
formed by Beethoven and Brahms, reflecting their heritage. Stravinsky
and Stockhausen forms another cluster and Mozart remains in isolation,
as like Bach and Monteverdi. Both relations were also present in the
planar space shown in Figure \ref{fig:pca}.

\section{Comparisons with Philosophers Analysis}

\begin{figure}
  \begin{center}
    \includegraphics[width=0.45\textwidth]{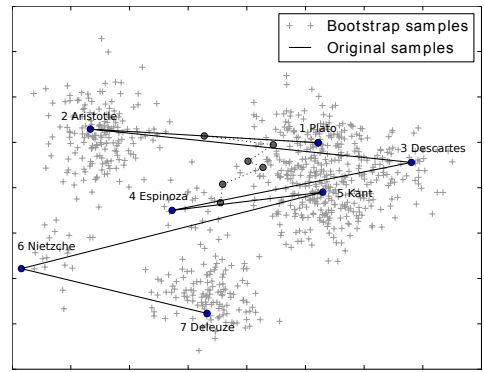}
  \end{center}
  \caption{\it 2-dimensional projected philosophical space.}
  \label{fig:phipca}
\end{figure}

The results of composers analysis are surprising when compared with philosophers~\cite{Fabbri}. 
It is important to note that we preserved the
number of characteristics and performed the same bootstrap method to
generate a larger set of samples, making possible this
comparison. The variances after PCA (Table \ref{tab:varphi}) concentrates in the four
first new axis, similar to the variances for composers shown at Table \ref{tab:tableC}. If we compare the discussed musical space
with the philosophical one in Figure \ref{fig:phipca} we
identify opposite movements along all the philosophy history in contrast
to music. This reveals a notorious characteristic of the way
philosophers seem to have evolved their ideas, driven by opposition ($W_{i,j}$), as shown in Table \ref{tab:tablephiOI}, while
composers tend to be more influenced by their predecessors as far as their dialectics measures are
concerned ($1/d_{i \rightarrow k}$).

In general, the musical movements had minor opposition and,
remembering the beginning of this work, it reflects the
master-apprentice
tradition present in music: the composers tend to build their own
works confirming their precursors legacy, resulting in a greater
dialectics than the philosophers related measures.
This reveals a crucial difference
considering the \textit{memory treatment} along the development of
philosophy and music: using the same techniques this article does~\cite{Fabbri},
we could verify that a philosopher was influenced by the
opposition of ideas from his direct predecessor, while here composers were commonly
influenced by their both predecessors. Therefore, we can argue that philosophy
presents a \textit{memory-1} state, while music presents
\textit{memory-2}, considering \textit{memory-N} being as number $N$
of past generations whose influence on a philosopher or
composer is being considered. Considering the linearity of musical movements we can
identify the abscissa as a ``time axis'' representing the development
of music along the history, with some composers
like Beethoven returning to Monteverdi and others advancing to the
modern age like Stravinsky and Stockhausen.

\begin{table}[ht]
\caption{\label{tab:varphi}New variances after PCA for philosophers
  scores in percentages.}

\begin{tabular}{|c||c|}
\hline
Eigenvalue  & Value     \\ \hline

$\lambda_1$ &  40 \% \\
$\lambda_2$ &  23 \% \\
$\lambda_3$ &  13 \% \\
$\lambda_4$ &  10 \% \\
$\lambda_5$ &   5 \% \\
$\lambda_6$ &   4 \% \\
$\lambda_7$ &   3 \% \\
$\lambda_8$ &   2 \% \\
\hline

\end{tabular}
\end{table}

The opposition and skewness indices for philosophers listed in Table
\ref{tab:tablephiOI} endorses the minor role of opposition in
composers at the period considered. We can observe strong opposition
in philosophical moves contrasted to small opposition in
musical movements. Also, the dialectics presents a
phase difference suggesting knownledge and aesthetics 
transfer latency between each of these human fields.

\begin{table}
\caption{\label{tab:tablephiOI}Opposition and skewness indices for each
of the six philosophical moves.}

\begin{tabular}{|c||c|c|}
\hline
Philosophical Move & $W_{i,j}$ & $s_{i,j}$ \\
\hline \hline
Plato $\rightarrow$ Aristotle     & 1.0    & 0 \\
Aristotle $\rightarrow$ Descartes & 0.8740 & 1.1205 \\
Descartes $\rightarrow$ Espinoza  & 0.9137 & 2.3856 \\
Espinoza $\rightarrow$ Kant       & 0.6014 & 1.6842 \\
Kant $\rightarrow$ Nietzsche      & 1.1102 & 2.9716 \\
Nietzsche $\rightarrow$ Deleuze   & 0.3584 & 2.4890 \\
\hline
\end{tabular}
\end{table}

\begin{figure}[ht]
        \begin{center}
                \includegraphics[width=0.45\textwidth]{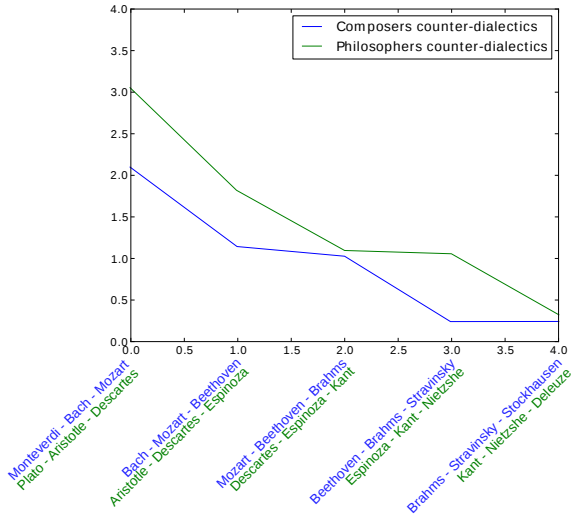}
        \end{center}
        \caption{\it Comparison between composers and philosophers
          counter-dialectics indices}
        \label{fig:comparingdialectics}
\end{figure}

When comparing dialectics, other curious facts arise: the dialectics
indices for musicians in Table \ref{tab:tableE} are considerably stronger moves than for
philosophers in Table \ref{tab:tablephiE}. Both indices are also shown in Figure
\ref{fig:comparingdialectics} where we can see a constant decrease
of counter-dialectics. This makes it possible to argue
that dialectics is stronger in music where a
constantly return to the origins are clearly visible. This reveals the nature of the
musical development, based on the search for a unity. Using the words
of Webern, the search for the ``comprehensibility'' but always
influenced by their old masters.

\begin{table}
\caption{\label{tab:tablephiE} Counter-dialectics index for each
of the five subsequent pairs of philosophical moves, considering all components.}

\begin{tabular}{|c||c|}
\hline
Philosophical Triple & $d_{i \rightarrow k}$ \\
\hline \hline
Plato $\rightarrow$ Aristotle $\rightarrow$ Descartes    & 3.0198 \\
Aristotle $\rightarrow$ Descartes $\rightarrow$ Espinoza & 1.8916 \\
Descartes $\rightarrow$ Espinoza $\rightarrow$ Kant      & 1.1536 \\
Espinoza $\rightarrow$ Kant $\rightarrow$ Nietzsche      & 1.1530 \\
Kant $\rightarrow$ Nietzsche $\rightarrow$ Deleuze       & 0.2705 \\
\hline
\end{tabular}
\end{table}

\section{Concluding Remarks}

Motivated by the understanding of how innovation evolves in music
history, we extended a quantitative method
recently applied to the study of philosophical
characteristics~\cite{Fabbri} and compared the results. Statistical
methods is nowadays commonly used for the study of music features and
composers productivity, but analysis of
composers characteristics modification along the music history has been less
explored. The method differs on the
aspect of how the characteristics concerning composers are treated:
scores are assigned to each feature common in musical
works. These scores reveal not the
exact profile of composers, but a tendency of how their
techniques are usually present. To make the simulation more
realistic, we considered not just the small number of 7 composers, but
derived other 1000 new ``artificial composers'' through a bootstrap
method. A larger data set made possible the statistical analysis,
considering not just the original scored composers, but other samples
that respect the historical presence of the formers. This other thousand
composers were modeled by a
probabilistic distribution, and avoided a biasing caused by
the use of only 7 composers.
In order to investigate the
relationship between this scorings we applied Pearson correlation
analysis. The results demonstrated a strong correlation between some
characteristics, which allows us to group this values, creating a
reduced number of features that summarizes the most important
characteristics. PCA was also applied to these components, reducing
the complex space to a planar graph where some of the most interesting
properties can be visualized. 

Historical landmarks in music are
well-defined in the planar space, like the isolation of Bach, Mozart
and Stockhausen, the
proximity between Beethoven and Brahms and the distance from Bach and Mozart, the heterogeneity of
Stravinsky and the vanguard of contemporary composers
like Stockhausen. Even not so visible relations, like the trend to return to the
maximum domain of polyphony -- present on Renaissance -- by Beethoven
could also be clearly observable, demonstrating the chronological nature of the
space. 

The dichotomy between
master-apprentice tradition on music and the quest for innovation that
opened this discussion could be visualized quantitatively. Each
composer demonstrated his own style, differing considerably from his
predecessor -- clearly shown when analyzing pairs of subsequent composers like
Bach and Mozart, Mozart and Beethoven or Stravinsky and
Stockhausen. Otherwise, the inheritance of predecessors styles is also
present when analyzing the direct relations between Mozart and
Beethoven or Beethoven and
Brahms, or indirect ones between Bach and Beethoven
or Beethoven and Monteverdi. The entire scenario presented
a ``continual pattern'' between
composers -- motivated by the influence of theirs predecessors -- but also showed a force
repelling both of them: the innovation, or in the words of William
Lovelock~\cite{Lovelock}, the ``experimentation'' that makes progress possible.

Along the analysis we noticed interesting differences when comparing
composers with philosophers. While on philosophy the
innovation is notably marked by opposition of each philosophers ideas,
it is less present for music composers. The lack of strong 
opposition movements and proeminent presence of dialectics in musical space indicates the music innovation is driven by
a constant heritage of each composer from his predecessors. We
represented this characteristic referring to a \textit{memory state}
where philosophers shows \textit{memory-1} -- each philosopher was
influenced by opposite ideas of its direct predecessor -- while
composers shows \textit{memory-2} -- inheriting the style of their
both direct predecessors.
The
analysis of both dialectics values also shown surprising
results: on philosophy the dialectics indices are arranged on a
increasing series -- showing a strong influence of
dialectics to philosophy development -- the dialectics indices on
music exhibits the same pattern, but with an offset. This behavior presumably indicates a
constant quest for coherence by the composers, a fact notably observed by
the studies of Anton Webern~\cite{Webern} should have somewhat the same
kernel and a lattency between the effects.

Another result is that the quantitative methodology initially applied to the analysis of philosophy~\cite{Fabbri}
proved to be extensible to other fields of knowledge -- in this case music --
reflecting with considerable efficiency details concerning the specific field. 

Computational analysis of music scores could be
applied to automate the quantification of composers characteristics, like
identification of melodic and harmonic patterns or the presence or not of
polyrhythms, motivic and harmonic stability~\cite{Correa}. More composers could be
inserted in the set for the analysis of a wider time-line, possibly
including more representatives of each music period.


We want to end this work going back to Webern,
who early envisioned these relations: ``It is clear that where relatedness and unity are omnipresent,
comprehensibility is also guaranteed. And all the rest is
dilettantism, nothing else, for all time, and always has been. That's
so not only in music but everywhere.''

\begin{acknowledgments}
Luciano da F. Costa thanks CNPq (308231/03-1) and FAPESP (05/00587-5)
for sponsorship. Gonzalo Travieso thanks CNPq (308118/2010-3) for sponsorship. 
Vilson Vieira and Renato Fabbri is grateful to CAPES and 
the Postgrad Committee of the IFSC.
\end{acknowledgments}

\nocite{*}
\bibliography{musimetrics}

\end{document}